# A Simple Elementary Proof of P=NP based on E. F. Codd's Relational Model[*]

Aizhong Li[†]

Corporola Inc.
18 Sunvale Way, Ottawa, Ontario, Canada K2G 6Y2

**ABSTRACT:** The P versus NP problem is studied under E. F. Codd's relational model. I found that the term "*complete configuration*" is unnecessary and harmful in computational complexity theory because of excessive symbol redundancy. For an input, its valid sequences of complete configurations are normalized into a relational model of shared *trichoices* with no redundancy. To simplify the problem, a polynomial time nondeterministic Turing machine is polynomially reduced to a *periodic machine*, which only reverses its tape head displacement at the tape ends. By enumerating all the $O(p(n))$ shared trichoices, a polynomial time $p(n)$ periodic machine is simulated in time $O(p^4(n))$ under logarithmic cost. A simple elementary proof of P=NP is obtained.

## 1. Introduction

**The P versus NP problem** is studied herein. It is the most important open problem in computer science. For its initial works, see S. A. Cook's seminal paper [5] and R. M. Karp's seminal paper [13]. For its surveys, see [3, 6, 9, 20]. It relates to *nondeterministic Turing machine* (choice machine by A. M. Turing) and their *computation*. Let us review these two core terms, under E. F. Codd's relational model [4], to find out why we all have failed to resolve the P versus NP problem.

### 1.1 Nondeterministic Turing Machine

Recall that a single-tape (infinite only to the right) **nondeterministic Turing machine** (NTM) can be denoted by $\mathbf{N} := (Q, Y, \Sigma, \delta, q_0, \vdash, b, F)$ after J. E. Hopcroft and J. D. Ullman [11], where

- Domain Q is a finite set of states. The initial state $q_0 \in Q$. The accepting states $F \subseteq Q$.
- Domain Y is a finite set of tape symbols. The input symbols $\Sigma \subset Y$. The blank symbol $b \in Y - \Sigma$. The "special" read-only left endmarker $\vdash \in Y - \Sigma - \{b\}$ for easy treatment herein.
- Finite relation $\delta \subseteq (Q-F) \times Y \times Q \times Y \times D$, where $\mathbf{D} := \{_L, _R\}$, with $_L := -1$ and $_R := +1$ after S. A. Cook [6], for one cell (or square) tape head displacement *left* or *right*. A **move** $(p, x, q, y, d) \in \delta$ consists of a current **condition** $(p, x)$ and a next **choice** or **machine operation** $(q, y, d)$. Let *next-choices function* $\boldsymbol{\delta(p, x)} := \{(q, y, d) / (p, x, q, y, d) \in \delta\}$. To avoid falling off the left-end, $\delta$ is subject to *the constraint* $\forall_{p \in Q} \delta(p, \vdash) \subseteq Q \times \{\vdash\} \times \{_R\}$.

If $\forall_{(p, x) \in (Q-F) \times Y} |\delta(p, x)| \leq 1$, NTM N degenerates to a **deterministic Turing machine** (DTM).

An NTM is a finite relation. It is natural to study its computations under E. F. Codd's relational model [4], in terms of polynomial time *deterministic* set operations extensively performed on relational databases [12, 19]. To do so, computations needs to be *normalized* into relation(s).

### 1.2 Normalized Computation

Originated from A. M. Turing's pioneering paper [18], in the literature, on an input, a valid *computation* (a *working* or a *behavior*) of an NTM N is described or defined as a valid sequence of complete configurations. For a time (complexity) $T(n)$ NTM N, on an input of length $n$, there are $O(T(n)|Y|^{T(n)})$ complete configurations, at worst.

A **complete configuration** is a triple $(\alpha, p, \beta)$ in $(Y^*) \times Q \times (Y^*)$. The interpretation of complete configuration $(\alpha, p, \beta)$ is that NTM N is in state $p$ with string $\alpha\beta$ on its tape from the left-end and with its tape head scanning the leftmost symbol of $\beta$ (or scanning $b$ if $\beta = \varepsilon$, the empty string).

---

[*] The algorithms in this paper are patent pending. See https://arxiv.org/abs/1809.04519 for its latest version.
[†] Email: AizhongLi@Gmail.com

Let $C$ and $C'$ be two compete configurations. $C'$ is ***a successor of*** $C$, denoted by $C \to C'$, if
- $C=(\alpha z, p, x\beta)$, $z \in Y$, $(q, y, \text{L}) \in \delta(p, x)$, and $C'=(\alpha, q, zy\beta)$; or  (1)
- $C=(\alpha, p, x\beta)$, $(q, y, \text{R}) \in \delta(p, x)$, and $C'=(\alpha y, q, \beta)$.  (2)

For an NTM N on input X, let the initial complete configuration $C_{-1} := (\varepsilon, q_0, \vdash X)$. A sequence (or time series) $C_0, C_1, ..., C_t$ of complete configurations is ***valid*** if $\forall_{i \in [0, t]}\, C_{i-1} \to C_i$.

In the literature, a time complexity $T(n)$ NTM is algorithmically simulated by searching for an accepting complete configuration in the space of $O(T(n)|Y|^{T(n)})$ successive complete configurations at worst. There is no trouble in using complete configurations in qualitative *computability* study [18]. However, in quantitative *computational complexity* study, every symbol is measured. The P versus NP problem is <u>*independent of*</u> "complete configuration".

- Assume P=NP. Let an NTM N be in polynomial time $p(n)$, then there are $O(p(n)|Y|^{p(n)})$ complete configurations, at worst. There is no DTM in polynomial time which can process all the $O(p(n)|Y|^{p(n)})$ complete configurations as wholes. "Something" more succinct than "complete configuration" must exist. As will be shown herein, "trichoice" is such one.
- Assume P≠NP. P≠NP cannot be obtained based on complete configurations exclusively, because they contain redundant symbols and are not optimal in describing computations. As will be shown herein, the relational model of trichoices is optimal without redundancy.

J. Hartmanis and R. E. Stearns' seminal paper [10] founded computational complexity theory. On pages 286-287 therein, they said clearly that "*The machine operation is our basic unit of time.*" Instead of $O(p(n)|Y|^{p(n)})$ lengthy complete configurations, there are only $|Q \times Y \times D|$ succinct machine operations or choices. Observe (1) and (2), complete configurations contain excessive redundancy. All the symbols but in the last scanned cell, $\alpha$, $z$, and $\beta$, are duplicated $|Q \times Y \times D|$ times nondeterministically at each unit of time, at worst. <u>Complete configurations are harmful or disastrous for computational complexity</u> because the duplicated symbols waste $O(|Y|^{p(n)})$ space and time. Fortunately, A. M. Turing said in [18], "*The changes of the machine and tape between successive complete configurations will be called the moves of the machine.*" By tracking successive changes only, a valid sequence of complete configurations can be normalized uniquely into a valid ***sequence of moves*** with less redundancy. Because every current *condition* can be obtained from $\vdash$, input X, $b$, the initial state, and/or at most two previous *choices*, a valid sequence of moves, in turn, can be normalized uniquely into a valid ***sequence of choices*** with no redundancy, as implied by J. E. Hopcroft and J. D. Ullman on pages 163-164 in [11]. In accordance with the basic unit of time complexity, from now on, <u>*a* (*valid normalized*) ***computation*** *is defined as a valid sequence of choices*</u>, with the unnecessary and harmful term "complete configuration" abolished.

On ***input*** $\mathbf{X} := \mathbf{X_1 X_2 \ldots X_n}$ in $\Sigma^n$, let $\mathbf{X_0} :\equiv\, \vdash$ and $\forall_{i > n}\, \mathbf{X_i} := b$. Initially, at epoch time 0, cell $i$ holds ***initial symbol*** $X_i$ starting from the leftmost cell 0 and M is in state $q_0$ reading $X_0$. For a choice $c$, let $c.q$, $c.y$, and $c.d$ denote its state, its symbol, and its tape head displacement. Formally, a (*valid normalized*) ***computation*** is such a valid time series $c_0, c_1, \ldots, c_t$ of choices that

- $c_0 \in \delta(q_0, X_0)$ at time 0 and tape head position **0**;
- $c_{j+1} \in \delta(c_j.q, c_{w(j)}.y)$ on reading the previously written symbol, where ***w(j)*** is the maximal time $i$ such that $i \leqslant j \wedge D(i, j)=0$; otherwise, $c_{j+1} \in \delta(c_j.q, X_{D(0, j)})$ on reading an *initial symbol*;

where $\mathbf{D(i, j)} := c_i.d + c_{i+1}.d + \ldots + c_j.d$, the sum of displacements from time $i$ to $j$. If $i > j$, $\mathbf{D(i, j)} := 0$.

## 1.3 Normalized Relational Model of Shared Trichoices for Computations

Observe the dependency among choices in valid normalized computations. A choice, $c_{j+1}$, depends on *at most* two previous choices, $c_j$ and $c_{w(j)}$. Following the database normalization procedure in [4] and best practices, it is not difficult to find out that, for an NTM N of time $T(n)$ on an input X, all its $O(|Q \times Y \times D|^{T(n)})$ valid sequences of choices collectively can be normalized (or decomposed, deserialized) into ***the relational model of $O(T^3(n))$ shared trichoices*** with no redundancy, where a ***trichoice*** consists of *at most three choices*, from a valid time series $c_0, c_1, \ldots, c_t$ of choices,

1. a *current choice* $c_{j+1}$, which is *executed* at time $j+1 \in [1, t]$ on cell ***h*** := D(0, *j*);
2. <u>a reference[‡] to its *predecessor choice* $c_j$</u> at time *j* on cell *h*+1 or *h*−1, for current state;
3. <u>a reference to its *writer choice* $c_{w(j)}$</u> at time *w(j)* on cell *h*, for currently read symbol.

A trichoice represents not only a move but also the ternary relationship among three choices, the fundamental inner structure of computations. Instead of $O(T(n)|Y|^{T(n)})$ lengthy complete configurations, there are at most $O(T^3(n))$ succinct trichoices because $0 \leq j, h, w(j) < T(n)$; and *j*+1, *h*+1, *h*−1, and *h* are functional dependents. In analogy with complete configurations, **the main idea of this paper** is to simulate an NTM N by enumerating all the $O(T^3(n))$ trichoices recursively. However, it is difficult to backtrack the *writer choice*(s) from a predecessor choice efficiently.

Fortunately, for the P versus NP problem, by polynomial time reductions between NTM's [5, 13], it is unnecessary to simulate all the NP NTM's. To avoid the difficult in writer choice backtracking without losing generality, *periodic machines* are coined. A ***periodic machine*** is an oblivious and polynomial bounded NTM, <u>which only reverses its tape head displacement at the ends of its tape</u>. Then, *w(j)* becomes a function of *j* and *n*. The *writer choice*(s) can be backtracked by trichoice deletions. The enumeration of trichoices is simplified and consequently the simulation of a periodic machine becomes efficient. Following this relational approach, the following main results are obtained.

### 1.4 The Main Results

A polynomial time NTM is polynomially reduced to a periodic machine. For a periodic machine M in polynomial time $O(n^k)$ with constant k, on an input of length *n*, all its valid sequences of choices are normalized into the relational model of $O(n^k)$ shared trichoices. By enumerating all the $O(n^k)$ trichoices, proved by mathematical induction, periodic machine M is simulated by an intuitive (deterministic) random-access machine program in time $O(n^{4k})$ under logarithmic cost. A simple elementary proof of P=NP is obtained.

The proof is based on the following previous works, mainly:
- A periodic machine is N. Pippenger and M. J. Fischer's oblivious machine [15], extends J. Myhill's *linear* bounded automaton (LBA) [14] to *polynomial* and M. Sipser's sweeping automaton [17] with writing ability. Periodic machines *simplify* the P versus NP problem.
- E. F. Codd's relational model [4] is ***<u>the basis</u>*** of computation representation and simulation.
- In the enumeration of trichoices collectively by recursion on time, a course-of-values recursive function [16] is extended to a course-of-values recursive relation. Most importantly, mathematical induction becomes *the major method of proof*.
- Following R. Fagin [7], A. V. Aho and J. D. Ullman [2]; finite sets, transitive closures, and least fixed points are used as *the key tools* for writer choice backtracking herein.
- Inspired by N. Immerman [12] and M. Y. Vardi [19] on polynomial time queries; herein, *in the reverse direction*, a polynomial time NTM is simulated by a polynomial time query.

Based on finite set theory, this paper is organized as follows:
- In Section 2, a periodic machine (PM) is defined formally.
- In Section 3, the basic properties of a PM are studied.
- In Section 4, the relational model of trichoices is established for a PM on an input.
- In Section 5, a polynomial time PM is simulated deterministically in polynomial time.

---

[‡] *References* (foreign keys [4], pointers, or <u>links</u>) make (long) subsequences shared without being duplicated, in analogy with citation references, like "[Turing, 1936]" or "[18]", in scientific papers. A reference and its referent may be the same by value. For example, in move (*p*, *x*, *q*, *y*, *d*)∈δ, current state *p* in the first column in δ, except $q_0$, is a foreign key reference to next state *p* in the third column. δ is *recursive* or *self-referencing*.

## 2. Periodic Machines

Throughout this paper, all numbers are non-negative integers. The logarithmic cost criterion in [1] is adopted for all random-access machine (RAM) programs, i.e., *deterministic computer* programs.

In order to make choices based on current state, symbol read, and last displacement, a periodic machine's states Q′⊆Q×D, i.e., last tape head displacement is treated as a part of a state, based on the Turing machine construction technique on page 153-154 in [11]. A periodic machine's states Q′ are partitioned into $Q_L$ and $Q_R$ by displacement L and displacement R, respectively. The initial state $q_{0L} \in Q_L$ holds the *imaginary* last displacement L, for easy treatment. Depicted in Fig. 1, a periodic machine is an oblivious, polynomial bounded, and sweeping NTM.

**Definition 1.** Formally, a single-tape ***periodic machine*** (PM) is denoted by the 10-tuple
  $M := (Q_L, Q_R, Y, \Sigma, \delta, q_{0L}, \vdash, b, \dashv, F)$, where

- The states $Q' := Q_L \cup Q_R$, $Q_L \cap Q_R = \phi$, and $q_0 \in Q_L$.
- $(Q', Y, \Sigma, \delta, q_{0L}, \vdash, b, F)$ constitutes an NTM N.
- The left endmarker $\vdash$ and the right endmarker $\dashv$ are "special" read-only symbols in $Y-\Sigma-\{b\}$.
- The relation $\delta \subseteq (Q'-F) \times Y \times Q' \times Y$. A *move* $(p, x, q, y) \in \delta$ is executed in the sense that machine M
  - changes its currently read symbol $x$ to $y$, first;
  - changes its current state $p$ to $q$. State change includes tape head displacement by L or R, depending on whether $q \in Q_L$ or $q \in Q_R$.

Let next-choices function $\delta(p, x) := \{(q, y) / (p, x, q, y) \in \delta\}$ for $(p, x) \in Q' \times Y$. It implies that $\forall_{(p, x) \in F \times Y} \delta(p, x) = \phi$. PM's relation $\delta$ is subject to **the constraint** below *to only reverse its tape head displacement at the ends of its tape* as depicted in Fig. 1, while keeping the two endmarkers distinct especially, where $I := Y - \{\vdash, \dashv\}$:

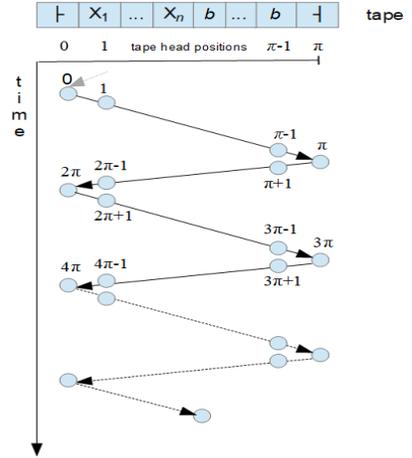

**Fig. 1.** The sweeps of the tape head

- $\forall_{p_L \in Q_L} \delta(p_L, \vdash) \subseteq Q_R \times \{\vdash\}$,   i.e., M reverses tape head displacement from L to R at the left-end only.
- $\forall_{p_R \in Q_R, x \in I} \delta(p_R, x) \subseteq Q_R \times I$,   i.e., M *sweeps* from the left-end to the right-end in the same displacement R.
- $\forall_{p_R \in Q_R} \delta(p_R, \dashv) \subseteq Q_L \times \{\dashv\}$, and i.e., M reverse tape head displacement from R to L at the right-end only.
- $\forall_{p_L \in Q_L, x \in I} \delta(p_L, x) \subseteq Q_L \times I$.   i.e., M *sweeps* from the right-end to the left-end in the same displacement L.

$\forall_{p_L \in Q_L}$ condition $(p_L, \dashv)$ is *unreachable*. $\forall_{p_R \in Q_R}$ condition $(p_R, \vdash)$ is *unreachable*.

On ***input*** $X := X_1 X_2 \ldots X_n$ in $\Sigma^n$, to allocate proper workspace, choose $\pi \in [n+1, n^{k'}+k']$ with a constant k′. Let $X_0 := \vdash$, $\forall_{i \in [n+1, \pi-1]} X_i := b$, and $X_\pi := \dashv$ by default. Confined to NP herein, a PM M is *polynomial bounded* in space $\pi+1 \leq n^{k'}+k'+1$. Herein $\pi := 2^m$ in $[n+1, n^{k'}+k']$ for easy treatment only. Initially, at time 0, cell $i$ holds *initial symbol* $X_i$ for $i \in [0, \pi]$ and M is in state $q_{0L}$ reading $X_0$. For the polynomial bound equivalence between $n^{k'}+k'$ in [6] and $O(n^k)$ in [1], see Theorem 12 in the Appendix.

## 3. The Basic Properties of Periodic Machines

Periodic machines are powerful enough to express all the NP-problems.

**Theorem 1.** For an NTM N in polynomial time $n^k+k$ after S. A. Cook [6], an equivalent PM M in polynomial time $O(n^{2k})$ can be constructed. (For its proof, see the Appendix.)

The three oblivious properties below simplify a PM's simulation. Let % be the remainder operator.

**Theorem 2.** For a PM M on input X at reachable time $t$, its *tape head position* is

$$h(t, n) = \begin{cases} t\%\pi & \text{if } t\%(2\pi) < \pi, \\ \pi - t\%\pi & \text{otherwise} \end{cases} \text{of fundamental period } 2\pi.$$

*Proof.* In Fig. 1, at reachable time *t*, there are two exclusive cases:
- in a sweep from the left-end (inclusive) to the right-end (exclusive), i.e., $t\%(2\pi) < \pi$: the tape head position is $t\%\pi$ from the left-end, the default reference point.
- in a sweep from the right-end (inclusive) to the left-end (exclusive): the tape head position is $t\%\pi$ from the right-end, i.e., $\pi-t\%\pi$ from the left-end. (The fundamental period is $2\pi$.) □

**Theorem 3.** For a PM M on input X in polynomial time $O(n^k)$ with constant k, *the last time,* when cell $h(t, n)$ was written before reachable time *t*, was

$$w(t, n) = \begin{cases} undefined & \text{if } 0 \le t \le \pi, \\ t-2\pi & \text{if } t \ge 2\pi \text{ and } t\%\pi = 0, \\ t-2(t\%\pi) & \text{otherwise.} \end{cases}$$

Moreover, $w(t, n)$ can be calculated by a RAM program in time $O(\log n)$ under logarithmic cost and $2 \le t - w(t, n) \le 2\pi$ when $w(t, n)$ is *defined*, i.e., when $t > \pi$.

*Proof.* In Fig. 1, there are three exclusive cases, depending on the value of *t*:
- $0 \le t \le \pi$: Cell $h(t, n)$ has never been written before time *t*, therefore $w(t, n)$ is *undefined*.
- $t \ge 2\pi$ and $t\%\pi = 0$, i.e., a tape end is reread: It took $2\pi$ steps. Therefore, $w(t, n) = t - 2\pi$.
- a middle cell is reread: It took $2(t\%\pi)$ steps, where $t\%\pi$ steps were taken from the last scanned tape end to cell $h(t, n)$, and vice versa. Therefore, $w(t, n) = t - 2(t\%\pi)$ in this case.

Because $\pi := 2^m$, $t\%\pi$ can be implemented in $\log t - \log \pi \le O(\log n)$ bit shifts because *reachable time* $t < O(n^k)$ entails $\log t \le O(\log n)$, and $\pi \le n^{k'} + k'$ entails $\log \pi \le O(\log n)$. In total, $w(t, n)$ takes time $O(\log n)$ under logarithmic cost by a RAM program [1], where **log i** $:= \lfloor \log_2 i \rfloor + 1$ if $i \ne 0$; otherwise, **log i** $:= 1$. It is not difficult to verify $2 \le t - w(t, n) \le 2\pi$ when $t > \pi$. □

**Theorem 4.** For a PM M on input X in polynomial time $O(n^k)$, *the next time,* when cell $h(t, n)$ will be reread after time *t*, will be $r(t, n) = t + 2(\pi - t\%\pi)$ if it is reachable. Moreover, $r(t, n)$ can be calculated by a RAM program in time $O(\log n)$ under logarithmic cost and $2 \le r(t, n) - t \le 2\pi$.

*Proof.* In Fig. 1, it will take $2(\pi - t\%\pi)$ steps to reread cell $h(t, n)$, where $\pi - t\%\pi$ steps will be taken from cell $h(t, n)$ to the following tape end, and vice versa. Therefore, $r(t, n) = t + 2(\pi - t\%\pi)$. Similar to Theorem 3, $r(t, n)$ can be calculated in time $O(\log n)$. It is easy to verify $2 \le r(t, n) - t \le 2\pi$. □

## 4. The Relational Model of Shared Trichoices for Computations

**Definition 2.** For a PM M on input X, *a (**valid normalized**) t-computation* (or computation, for short) is such a valid time series $c_0, c_1, \ldots, c_t$ of choices that $\forall_{i \in [0, t]}$
- choice $c_0 \in \delta(q_{0L}, X_0)$ is *executed* at time $i = 0$,
- choice $c_i \in \delta(c_{i-1}.q, X_i)$ is *executed* at time $i \in [1, \pi]$,
- choice $c_i \in \delta(c_{i-1}.q, c_{w(i, n)}.y)$ is *executed* at time $i > \pi$, (Recall that $w(i, n) \le i - 2$ by Theorem 3.)

where $c.q$ and $c.y$ denote the state and the symbol of a choice $c \in Q' \times Y$, respectively.

If $c_0, c_1, \ldots, c_t$ is a *t*-computation, then $\forall_{i < t} c_0, c_1, \ldots, c_i$ is an *i*-computation. In a *t*-computation, at most three choices are required to represent a move and direct relationships among choices.

**Definition 3.** For a *t*-computation $c_0, c_1, \ldots, c_t$ of M on input X, its *i-th trichoice* ($\underline{c_{w(i, n)}}, \underline{c_{i-1}}, c_i$), for time $i \le t$, is interpreted, under E. F. Codd's relational model [4], as follows:
- primary key choice $c_i$ is currently executed in this trichoice at time *i*. Choice $c_i \ne null$.
- foreign key choice $\underline{c_{i-1}}$ is a reference to primary key choice $c_{i-1}$ previously *executed* in trichoice ($\underline{c_{w(i-1, n)}}, \underline{c_{i-2}}, c_{i-1}$) if $i \ne 0$; otherwise, $\underline{c_{i-1}} := null$. Choices $c_{i-1}$ and $c_i$ have **predecessor-successor** relationship. At time *i*, current state $p_i := c_{i-1}.q$ if $i \ne 0$; otherwise, $p_i := q_{0L}$.

- foreign key choice $\underline{c_{w(i,n)}}$ is a reference to primary key choice $c_{w(i,n)}$ previously *executed* in trichoice ($\underline{c_{w(w(i,n),n)}}$, $\underline{c_{w(i,n)-1}}$, $c_{w(i,n)}$) if $i \notin [0, \pi]$; otherwise, $\underline{c_{w(i,n)}} := null$. Choices $c_{w(i,n)}$ and $c_i$ have **writer-reader** relationship, which is a transitive resultant of successive predecessor-successor relationships, critical in writer backtracking herein. At time $i$, currently read symbol $x_i := c_{w(i,n)}.y$ if $i \notin [0, \pi]$; otherwise, $x_i := X_i$.

Choice $c_i \in \delta(p_i, x_i)$ by the definition of *t*-computation $c_0, c_1, \ldots, c_t$. A choice is a writer reference, a predecessor reference, or a currently executed choice, depending on its position in a trichoice, an ordered triple. From now on, following the convention in [4], for a choice $c$, its reference $\underline{c}$ is *written* as $c$ itself with the underline omitted in a trichoice, i.e., *reference by value*.

Observe all the *t*-computations for a PM M on input X. If the number of *t*-computations exceeds $|\{null\}^2 \times (Q' \times Y) \cup \{null\} \times (Q' \times Y)^2 \cup (Q' \times Y)^3|$ at time *t*, by the pigeonhole principle, there are shared trichoices at each time. It is redundant to store the shared trichoices more than once. The "trick" is to store trichoices uniquely in **relational model R** as follows to save both space and time.

**Definition 4.** For a PM M on input X in polynomial time $O(n^k)$, its computations are ***decomposed*** (or normalized) into the time series $\mathbf{R} := R_0, R_1, \ldots, R_{O(n^k)-1}$ of *finite relations*, where, for $t<O(n^k)$,

$\mathbf{R}_t := \{(c_{w(t,n)}, c_{t-1}, c_t) \,/\, (c_{w(t,n)}, c_{t-1}, c_t)$ is the *t*-th trichoice of *t*-computation $c_0, c_1, \ldots, c_t$ of M on input X$\}$.

For a trichoice $(w, p, c) \in R_t$, its choices' ***occurrence times*** are $w(t, n)$, $t-1$, and $t$, respectively. $|R_t| \leq |\{null\}^2 \times (Q' \times Y) \cup \{null\} \times (Q' \times Y)^2 \cup (Q' \times Y)^3| = O(1)$, a constant upper bound. Invariant to input length *n*, sets Q′ and Y are treated as *constants*, because space complexity and time complexity are measured in input length *n*, conventionally [1]. By definition, $R_0 = \underline{\{(null, null, c_0) \,/\, c_0 \in \delta(q_{0L}, X_0)\}}$.

**Definition 5.** For a PM M on input X in polynomial time $O(n^k)$, let a time series $\Delta := \Delta_0, \Delta_1, \ldots, \Delta_t$ of subrelations satisfying $\forall_{i \in [0, t]} \Delta_i \subseteq R_i$. A *t*-computation $c_0, c_1, \ldots, c_t$ is called ***being composed*** (or denormalized) from subrelations $\Delta$ if $\forall_{i \in [0, t]} (c_{w(i,n)}, c_{i-1}, c_i) \in \Delta_i$.

If $\exists_{i \in [0, t]} \Delta_i = \phi$, no *t*-computation can be composed from $\Delta$. What if $\forall_{i \in [0, t]} \Delta_i = R_i$ at extreme?

**Theorem 5.** For a PM M on input X in polynomial time $O(n^k)$, the computations defined by Definition 2 are exactly the computations composed from R by Definition 5.

*Proof.* $\Leftarrow$ A *t*-computation composed from R by Definition 5 is definitely a *t*-computation.
$\Rightarrow$ If $c_0, c_1, \ldots, c_t$ is a *t*-computation by Definition 2, then $\forall_{i \in [0, t]} (c_{w(i,n)}, c_{i-1}, c_i) \in R_i$ because $c_0, c_1, \ldots, c_i$ is an *i*-computation. Therefore, *t*-computation $c_0, c_1, \ldots, c_t$ is composed from R. □

Relations R constitute a polynomial space $O(n^k)$ representation of $O(|Q' \times Y|^{n^k})$ computations of a PM in time $O(n^k)$. Most importantly, relations R yield the efficient simulation of a PM as follows.

## 5. The Efficient Simulation of a Periodic Machine

It is evident that PM M on input X can be simulated by calculating relations R. However, it is inefficient to calculate $R_t$ by Definition 4 literally. Instead, its equivalent $S_t$, by membership, will be calculated efficiently by a ***course-of-values recursive relation*** on time *t*, extending a course-of-values recursive function [16] as follows, where a *finite relation* acts as a Gödel number [8]:
- At time $t=0$, constant *finite relation* $\mathbf{S_0} := \underline{\{(null, null, c_0) \,/\, c_0 \in \delta(q_{0L}, X_0)\}}$ herein.
- *Finite relation* $S_{t+1}$ is calculated from previously calculated *finite relations* $S_0, S_1, \ldots, S_t$ by set operations, in analogy with arithmetic operations.

**Lemma 1.** For a PM M, $S_0 = R_0 = \underline{\{(null, null, c_0) \,/\, c_0 \in \delta(q_{0L}, X_0)\}}$ and $S_0$ can definitely be calculated in time $O(1)$ by a RAM program, independent of input. (Recall that $X_0 := \vdash$.) □

Inductively hypothesize that finite relations $S_0, S_1, \ldots, S_t$ have been calculated and $\forall_{i\in[0,\,t]} S_i = R_i$. If $S_t = R_t = \phi$, finite relation $S_{t+1} = R_{t+1} = \phi$; otherwise, $S_{t+1}$ is calculated from $S_0, S_1, \ldots, S_t$ as follows.

**Pivot** an **executed choice** $c_t$ from $S_t$, i.e., $\exists_{w,\,p}\,(w, p, c_t) \in S_t$, by **initializing** a time series $\Delta(c_t) := \Delta_0(c_t), \Delta_1(c_t), \ldots, \Delta_t(c_t)$ of subrelations to make choice $c_t$ unique at time $t$, where

---
$\Delta_i(c_t) := S_i$ for time $i < t$;
$\Delta_t(c_t) := S_t - \{(w, p, c) \,/\, (w, p, c) \in S_t \land c \neq c_t\}$.      i.e., $\Delta_t(c_t) := \{(w, p, c_t) \,/\, (w, p, c_t) \in S_t\}$.

---

**Fig. 2.** The initialization of subrelations $\Delta(c_t)$

After pivoting, M will be in state $c_t.q$ at time $t+1$. What is left is to find out which symbol(s) will be read by M in state $c_t.q$ at time $t+1$. There are two cases, depending on the value of time $t+1$:
1. if $t+1 \in [1, \pi]$, then symbol $X_{t+1}$ will be the only symbol to be read at time $t+1$; otherwise,
2. the symbol(s) to be read at time $t+1$ must be previously written at time $w(t+1, n) \leq t-1$ by Theorem 3. The crux is how to backtrack the writer(s) from predecessor $c_t$ inside $\Delta(c_t)$.

In general, let a time series $\Delta := \Delta_0, \Delta_1, \ldots, \Delta_t$ of subrelations satisfying $\forall_{i\in[0,\,t]} \Delta_i \subseteq S_i$.

Recall that a writer-reader relationship is a transitive resultant of successive predecessor-successor relationships. A valid reader must be in the transitive closure of its writer by predecessor-successor relationships. For an executed choice $c_i$ (enclosed) in $\Delta_i$, let its transitive closure $\lambda_\Delta^m(c_i)$ be executed choices in $\Delta_{i+m}$, by $m$ successive predecessor-successor relationships in $\Delta$. Formally,
- $\lambda_\Delta^0(c_i) := \{c_i \,/\, (w, p, c_i) \in \Delta_i\}$;      (note: the subscript $i$ of $c_i$ is used as a parameter.)
- $\lambda_\Delta^{m+1}(c_i) := \{c_{i+m+1} \,/\, c_{i+m} \in \lambda_\Delta^m(c_i) \land (w, c_{i+m}, c_{i+m+1}) \in \Delta_{i+m+1}\}$.

For $j \in [0, t]$, a trichoice $(c_{w(j, n)}, c_{j-1}, c_j) \in \Delta_j$ is **ill-referenced** with respect to $\Delta$, if
I.    $j > 0$ and $c_{j-1}$ is not an executed predecessor of $c_j$, $c_j \notin \lambda_\Delta^1(c_{j-1})$; or
II.    $j < t$ and $c_j$ has no executed successor, $\lambda_\Delta^1(c_j) = \phi$; or
III.   $j > \pi$ and $c_j$ is not in transitive closure of writer $c_{w(j, n)}$, $c_j \notin \lambda_\Delta^{j-w(j,n)}(c_{w(j, n)})$; or
IV.   $r(j, n) \leq t$ and $c_j$ has no reader in its transitive closure, $\neg(\exists c \in \lambda_\Delta^{r(j,n)-j}(c_j) \land (c_j, p, c) \in \Delta_{r(j, n)})$.

Ill-referenced trichoices are not executed in any $t$-computation composed from $\Delta$ and are subject to deletion to maintain referential integrity.

**Function $lfp(\Delta)$** in Fig. 3 is given to deletes all the ill-referenced trichoices in $\Delta$ repeatedly until the least fixed point (LFP) is reached for a polynomial time $O(n^k)$ PM M on X at time $t$.

In Fig. 3, at reachable time $j \leq t < O(n^k)$, by Theorem 3 and 4, both $w(j, n)$ and $r(j, n)$ can be calculated in time $O(\log n)$. By Theorem 3 and 4, $j - w(j, n) \leq 2\pi$ and $r(j, n) - j \leq 2\pi$. It is obtained that both $\lambda_\Delta^{j-w(j,n)}(c_{w(j, n)})$ and $\lambda_\Delta^{r(j,n)-j}(c_j)$ can be calculated in time $O(\pi) \geq O(n)$. Assume $O(n^k) \geq \pi$; otherwise, with written symbols never being read, PM M degenerates to a nondeterministic finite automaton (NFA) [11] after trapping all accepting states. By Theorem 2.1 in [11], an equivalent deterministic finite automaton (DFA) for an NFA can be constructed. Both the NFA and the DFA are of time complexity $O(n)$. Because $\forall_{j \leq t} |\Delta_j| \leq |S_j| = |R_j| = O(1)$, the set operations on $\Delta_j$ can all be finished in time $O(1)$. The "*for each*" loop repeats $O(1)$ times. The "*for time*" loop repeats $t+1$ times. The "*while*" loop repeats at most $O(n^k)$ times because it repeats after the first time only if the flag *isLFP* is reset to *false*, i.e., one trichoice is deleted. Therefore, $lfp(\Delta)$ is within time $O(n^{3k})$.

**Lemma LFP.** For a PM M on X in polynomial time $O(n^k)$, subrelations $lfp(\Delta)$ include exactly all the trichoices executed in the $t$-computations composed from $\Delta$. (See the Appendix, for its proof.) Moreover, as a RAM program, function $lfp(\Delta)$ takes time $O(n^{3k})$ under logarithmic cost.   □

*isLFP* := *false*;

*while* ¬*isLFP* do
  *isLFP* := *true*;
  *for* time $j := 0; j \leq t; j := j+1$
    *for each* trichoice $(c_{w(j, n)}, c_{j-1}, c_j) \in \Delta_j$
      *if* $((j>0 \wedge c_j \notin \lambda_\Delta^1(c_{j-1}))$      ∨      // Case I
        $(j<t \wedge \lambda_\Delta^1(c_j)=\phi)$      ∨      // Case II
        $(j>\pi \wedge c_j \notin \lambda_\Delta^{j-w(j,n)}(c_{w(j, n)}))$      ∨      // Case III
        $(r(j, n) \leq t \wedge \neg(\exists c \in \lambda_\Delta^{r(j,n)-j}(c_j) \wedge (c_j, p, c) \in \Delta_{r(j, n)})))$      // Case IV
      *then*
        *isLFP* := *false*; and $\Delta_j := \Delta_j - \{(c_{w(j, n)}, c_{j-1}, c_j)\}$;
*return* $\Delta$.

**Fig. 3.** Function *lfp*($\Delta$)

Every *t*-computation composed from $lfp(\Delta(c_t))$ must end at pivoted choice $c_t$. By Lemma LFP, subrelations $lfp(\Delta(c_t))$ include <u>exactly</u> all the trichoices executed in the *t*-computations composed from $\Delta(c_t)$. At time $t+1 \notin [1, \pi]$, PM M will be in state $c_t.q$ reading <u>exactly</u> the symbol(s) written by the executed choice(s) (enclosed) in ***lfp**_{w(t+1, n)}(\Delta(c_t))*, which is $\Delta_{w(t+1, n)}(c_t)$ returned by function $lfp(\Delta(c_t))$. Therefore, **relation** $S_{t+1}$, equal to $R_{t+1}$, both equal to all the $(t+1)$-th trichoices of all the $(t+1)$-computations, is calculated in Fig. 4, by pivoting each executed choice in $S_t$.

$S_{t+1} := \phi$;
*for each* $c_t \in \{c / (w', p', c) \in S_t\}$
  *if* $t+1 \in [1, \pi]$ *then*                                                                        Case 1
    $S_{t+1} := S_{t+1} \cup \{(null, c_t, c_{t+1}) / c_{t+1} \in \delta(c_t.q, X_{t+1})\}$;
  *else*                                                                                                       Case 2
    $S_{t+1} := S_{t+1} \cup \{(w, c_t, c_{t+1}) / (w'', p'', w) \in lfp_{w(t+1, n)}(\Delta(c_t)) \wedge c_{t+1} \in \delta(c_t.q, w.y)\}$.

**Fig. 4.** Calculation of $S_{t+1}$

In Fig. 4, at reachable time $t+1 < O(n^k)$, by Theorem 3, $w(t+1, n)$ can be calculated in time $O(\log n)$. $S_t$ has size $O(1)$. The loop repeats at most $|Q' \times Y| = O(1)$ times. In the loop, both $\delta(c_t.q, X_{t+1})$ and subrelation $lfp_{w(t+1, n)}(\Delta(c_t))$ has size $O(1)$. Therefore, including the initialization of $\Delta(c_t)$ in Fig. 2, $S_{t+1}$ can be calculated in time $O(n^{3k})$, mainly due to the time of $lfp_{w(t+1, n)}(\Delta(c_t))$ by Lemma LFP. Therefore, the following lemma is obtained.

**Lemma 2.** For a PM M on input X in polynomial time $O(n^k)$ at reachable time $t+1$, relation $S_{t+1}$, equal to $R_{t+1}$, can be calculated from $S_0, S_1, \ldots, S_t$ by a RAM program in time $O(n^{3k})$ under logarithmic cost. □

Based on Lemma 1 and 2, by mathematical induction on time *t*, the theorem below is obtained.

**Theorem 6.** For a PM M on input X in polynomial time $O(n^k)$ at reachable time *t*, relation $S_t$, equal to $R_t$, can be calculated in time $O(n^{3k})$ under logarithmic cost by a RAM program. □

**Theorem 7.** Whether a PM M in polynomial time $O(n^k)$ accepts input $X_1 X_2 \ldots X_n$ can be determined in time $O(n^{4k})$ under logarithmic cost by a (deterministic) RAM program.

*Proof.* As depicted in Fig. 5, in addition to checking whether $q_{0L} \in F$, by checking for an accepting state in each executed choice of each trichoice in relations $S_0, S_1, \ldots, S_{O(n^k)-1}$, whether PM M accepts input $X_1 X_2 \ldots X_n$ can be determined in time $O(n^{4k})$, mainly due to the calculations of the $O(n^k)$ relations $S_0, S_1, \ldots, S_{O(n^k)-1}$, each of which takes time $O(n^{3k})$ by Theorem 6. □

```
if q₀ₗ∈F then return true;
for time t := 0; t < O(nᵏ); t := t+1
    calculate Sₜ;
    if ∃₍w, p, c₎∈Sₜ c.q∈F then
        return true;
return false.
```

**Fig. 5.** The simulation of PM M

**Theorem 8** (Theorem 1.4 in [1]). The RAM under logarithmic cost criterion and the *multitape* DTM are polynomially related computational models.

**Theorem 9** (Lemma 10.1 and its Corollary 1 in [1]). An equivalent *single-tape* NTM in time $O(T^2(n))$ can be constructed for a time complexity $T(n)$ *multitape* NTM. An equivalent *single-tape* DTM in time $O(T^2(n))$ can be constructed for a time complexity $T(n)$ *multitape* DTM.

**Theorem 10.** For a periodic machine M in polynomial time, an equivalent single-tape DTM in polynomial time can be constructed.

*Proof.* For a PM M in polynomial time $O(n^k)$, by Theorem 7, an equivalent (deterministic) RAM program in polynomial time $O(n^{4k})$ under logarithmic cost is constructed. By Theorem 8, an equivalent *multitape* DTM in polynomial time can be constructed. By Theorem 9, in turn, an equivalent *single-tape* DTM in polynomial time can be constructed. □

**Theorem 11. P=NP.**

*Proof.* By Theorem 9, without losing generality, only single-tape Turing machines are considered for both P and NP. Moreover, by the proof of Theorem 7.1 in [11], only one-way infinite tape Turing machines are considered. That is, only NTM's defined in Section 1 are considered.

For a single-tape NTM N in polynomial time $n^k+k$ after S. A. Cook [6], by Theorem 1, an equivalent PM M in polynomial time $O(n^{2k})$ is constructed. By Theorem 10, an equivalent single-tape DTM in polynomial time is constructed. Therefore, NP⊆P.

By the definitions of P and NP, P⊆NP.

Because P⊆NP and NP⊆P, it is obtained that P=NP. □

## 6. Conclusions

Periodic machines simplify the P versus NP problem. To eliminate symbol redundancy, a sequence of complete configurations is normalized into a sequence of choices. By the dependency analysis of choices for a periodic machine in polynomial time $O(n^k)$, on an input of length $n$, its $O(|Q'\times Y|^{n^k})$ valid sequences of choices are collectively normalized into the relational model of $O(n^k)$ shared trichoices with no redundancy. By enumerating all the $O(n^k)$ trichoices, a time $O(n^{4k})$ random-access machine program under logarithmic cost criterion is obtained to simulate the periodic machine. P=NP is proved constructively. Trichoices reveal the inner structure of computations. E. F. Codd's relational model is not only a data model but also a computational model.

In applications, for NP problems, their computational time and space are reduced exponentially. Nondeterministic computation can be simulated efficiently on an existing (deterministic) computer.

## Appendix

**Theorem 1.** For an NTM N in polynomial time $n^k+k$ after S. A. Cook [6], an equivalent PM M in polynomial time $O(n^{2k})$ can be constructed.

*Proof.* As defined in Section 1, let NTM $N := (Q, Y, \Sigma, \delta, q_0, \vdash, b, F)$ in polynomial time $n^k+k$.

A PM $M := (Q_L, Q_R, Y_M, \Sigma, \delta_M, (q_0, L), \vdash, b, \dashv, F \times \{L, R\})$ is constructed to simulate N as follows, where $\dashv \notin Y$. Let input $X_1X_2...X_n$ be padded with blank symbols to make M and N share the initial tape layout $\vdash X_1X_2...X_n b^{\pi-n-1} \dashv$ when $k \neq 0$, where $\pi := 2^m$ if $n^k+k = 2^m$; otherwise, $n^k+k = 2^m+a$ with $0 < a < 2^m$, $\pi := 2^{m+1} < 2(n^k+k)$. By Theorem 12, there exists constant $k'$ such that $\pi < 2(n^k+k) \leq n^{k'}+k'$. $\pi$ is so big that NTM N must halt before ever reading the right endmarker $\dashv$. When $k=0$, the input $X_1X_2...X_n$ is not read, let the initial tape layout be $\vdash\dashv$, $\pi := 2^0 = 0^1 +1$. A **normal state** of PM M is $(p, d) \in Q \times D$, written as $p_d$, holding a combination of N's state $p$ and its associated tape head displacement $d$. Initially, NTM N is in state $q_0$ with the imaginary last head displacement $L$. PM M is in state $p_d$ if and only if NTM N is in state $p$ just after N has displaced tape head by $d$.

As depicted in Fig. 6, at the left-end or on the same displacement, PM M follows N one move by one move while tracking N's last tape head displacement; otherwise, when N reveres its tape head displacement, in between the tape ends, from $d$ to its reverse $\check{d} := -d$, in a move, $m$,

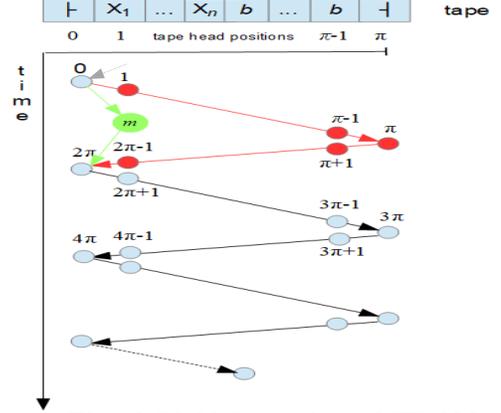

- M **records** previous displacement $d$ and move $m$ by entering new **compound state** $d_m$ and writing move $m$ as a new **compound symbol** on the tape to mark the current tape head position;
- M **skips** to the following tape end in state $d_m$ in last displacement $d$;
- M **changes** state from $d_m$ to $\check{d}_m$ at the tape end;
- M **skips** backwards in state $\check{d}_m$ in displacement $\check{d}$ to find compound symbol $m$ on the tape; and then
- M *"executes"* move $m$'s choice.

Compound states, $d_m$ and $\check{d}_m$, are so uniquely arranged as to avoid interference by any other move of N.

**Fig. 6.** PM M simulates NTM N

The components of PM M are defined formally, where $\Gamma := Q \times Y \times Q \times Y$ and $I := Y - \{\vdash\}$:

- $Q_L := Q \times \{L\} \cup \{L_m / m \in Q \times Y \times Q \times Y \times D\}$, i.e., **normal** and **compound states** for left sweeping.
- $Q_R := Q \times \{R\} \cup \{R_m / m \in Q \times Y \times Q \times Y \times D\}$, i.e., **normal** and **compound states** for right sweeping.
- $Y_M := (Y \cup \{\dashv\}) \cup Q \times Y \times Q \times Y \times D$, i.e., **simple symbols** and **compound symbols** are disjointed.
  A move $(p, x, q, y, d) \in Q \times Y \times Q \times Y \times D$ is written as $pxqyd$ for conciseness.
- $\delta_M$ is defined as follows: There are two cases, depending on N's last head displacement:

1. L (i.e., PM M is sweeping *left*, including its initial head displacement L):
    - $\forall_{p \in Q} \delta_M(p_L, \vdash) := \{(q_R, \vdash) / (q, \vdash, R) \in \delta(p, \vdash)\}$    i.e., M follows N at the left-end by their constraints.
    - $\forall_{p \in Q, x \in I} \delta_M(p_L, x) := \{(q_L, y) / (q, y, L) \in \delta(p, x)\}$   i.e., M follows N if no displacement change.
      $\cup \{(L_{pxqyR}, pxqyR) / (q, y, R) \in \delta(p, x)\}$. i.e., N reverses displacement between ends.
    - $\forall_{pxqy \in \Gamma, a \in I} \delta_M(L_{pxqyR}, a) := \{(L_{pxqyR}, a)\}$.    i.e., M skips to the left-end.
    - $\forall_{pxqy \in \Gamma} \delta_M(L_{pxqyR}, \vdash) := \{(R_{pxqyR}, \vdash)\}$.    i.e., M reverses displacement at the left-end.
    - $\forall_{pxqy \in \Gamma, a \in I} \delta_M(R_{pxqyR}, a) := \{(R_{pxqyR}, a)\}$.    i.e., M skips right until compound symbol $pxqyR$.
    - $\forall_{pxqy \in \Gamma} \delta_M(R_{pxqyR}, pxqyR) := \{(q_R, y)\}$.    i.e., M "executes" N's choice of move $(p, x, q, y, R)$.

2. L (i.e., PM M is sweeping *right*.):
    - $\forall_{p \in Q, x \in I} \delta_M(p_R, x) := \{(q_R, y) / (q, y, R) \in \delta(p, x)\}$.    i.e., M follows N if no displacement change.
      $\cup \{(R_{pxqyL}, pxqyL) / (q, y, L) \in \delta(p, x)\}$. i.e., N reverses displacement between ends.
    - $\forall_{pxqy \in \Gamma, a \in I} \delta_M(R_{pxqyL}, a) := \{(R_{pxqyL}, a)\}$.    i.e., M skips to the right-end.
    - $\forall_{pxqy \in \Gamma} \delta_M(R_{pxqyL}, \dashv) := \{(L_{pxqyL}, \dashv)\}$.    i.e., M reverses displacement at the right-end.
    - $\forall_{pxqy \in \Gamma, a \in I} \delta_2(L_{pxqyL}, a) := \{(L_{pxqyL}, a)\}$.    i.e., M skips left until compound symbol $pxqyL$.
    - $\forall_{pxqy \in \Gamma} \delta_M(L_{pxqyL}, pxqyL) := \{(q_L, y)\}$.    i.e., M "executes" N's choice of move $(q, x, q, y, L)$.

It is straight forward to convert next-choices function $\delta_M(p, x)$ into relation $\delta_M$ formally.

One move of N is simulated by PM M using at most $\pi+(\pi-1) < 4\times(n^k+k)-1$ moves if N reverses its tape head displacement at the second cell from the left-end, as depicted in Fig. 6. Therefore, the time complexity of PM M is in $(4\times(n^k+k)-1)\times(n^k+k) \leq O(n^{2k})$, including k=0, with $0^0 =1$. □

**Lemma LFP.** For a PM M on X in polynomial time $O(n^k)$ with constant k, subrelations $lfp(\Delta)$ include <u>exactly</u> all the trichoices executed in the $t$-computations composed from $\Delta$.

*Proof.* ⇐ In terms of Case I, II, III, and IV in Fig. 3, no trichoice executed in a $t$-computation composed from $\Delta$ is deleted by function $lfp(\Delta)$. Therefore, subrelations $lfp(\Delta)$ include at least all the trichoices executed in the $t$-computations composed from $\Delta$.

⇒ What is left is to prove that every trichoice ***in*** $lfp(\Delta)$, i.e., in $lfp_0(\Delta)$, or $lfp_1(\Delta)$, …, or $lfp_t(\Delta)$, is executed in at least one $t$-computation composed from $\Delta$. For easy treatment, this part of proof is restricted to such a ***standard*** $\Delta$ that for all $j\in[0, t]$ and for all trichoice $(c_{w(j, n)}, c_{j-1}, c_j)\in\Delta_j$, it satisfies

    A. $c_j\in\lambda_\Delta^1(c_{j-1})$, except $j=0$, and      i.e. $c_j$ has at least one executed predecessor including $c_{j-1}$ if $j\neq 0$.
    B. $\lambda_\Delta^1(c_j)\neq\phi$, except $j=t$;      i.e. $c_j$ has at least one executed successor if $j\neq t$.

by assuming that $\Delta$ is the least fixed point after iterative Case I and Case II deletions. If $\exists_{i\in[0, t]}$ $\Delta_i=\phi$, then $\forall_{i\in[0, t]} \Delta_i=\phi$ by Case I and Case II deletions. If so, it is trivially true that every trichoice in subrelations $lfp(\Delta)$, all empties, is executed in at least one $t$-computation composed from $\Delta$. In what follows, $\forall_{i\in[0, t]}\Delta_i\neq\phi$ is assumed.

This part of proof is obtained by mathematical induction on ***the size of*** $\Delta$, $\|\Delta\| := \sum_{i=0}^t \|\Delta_i\|$, where $\|\Delta_i\| := |\{c_i / (w, p, c_i)\in\Delta_i\}| -1$. Because $\forall_{i\in[0, t]}\Delta_i\neq\phi$, $\|\Delta\|$ is a sum of non-negative integers.

*Base case* ($\|\Delta\|=0$): If $\|\Delta\|=0$, then $\forall_{i\in[0, t]} |\{c_i / (w, p, c_i)\in\Delta_i\}|=1$, i.e., there is only one executed choice at each time. Let the executed choices be $\mathbf{c_0}, \mathbf{c_1}, …, \mathbf{c_t}$ enclosed in $\Delta_0, \Delta_1, …, \Delta_t$, respectively.

If $\exists_{i\in[0, t]} lfp_i(\Delta)=\phi$, then $\forall_{i\in[0, t]} flp_i(\Delta)=\phi$ by Case I and Case II deletions. In this case, it is trivially true that every trichoice in subrelations $lfp(\Delta)$, all empties, is executed in at least one $t$-computation composed from $\Delta$. In what follows, $\forall_{i\in[0, t]} lfp_i(\Delta)\neq\phi$ is assumed. Observe $lfp_i(\Delta)$.

- At time $i=0$, it is obtained that $lfp_0(\Delta)=\{(null, null, c_0)\}$, a singleton, because both $c_0$'s predecessor and $c_0$'s writer must be *null* by Definition 3.
- Observe predecessor-successor relationships when $i>0$. For any trichoice $(w, p_0, c_1)\in lfp_1(\Delta)$, choice $c_1$'s writer $w$ must be *null* by Definition 3. If $\exists_{p_0\neq c_0} \mathbf{(null, p_0, c_1)\in lfp_1(\Delta)}$, then $\lambda_\Delta^0(p_0)=\phi$ and $c_1\notin\lambda_\Delta^1(p_0)=\phi$; choice $p_0$ is not an executed predecessor of $c_1$ and $(null, p_0, c_1)$ must have been deleted from $lfp_1(\Delta)$ by Case I. If $\forall_{(null, p_0, c_1)\in lfp_1(\Delta)} p_0\neq c_0$, then $(null, null, c_0)$ must have been deleted from $lfp_0(\Delta)$ by Case II of no successor. Both lead to the contradiction that $lfp(\Delta)$ is not a least fixed point. Therefore, $p_0=c_0$ and $lfp_1(\Delta)=\{(null, c_0, c_1)\}$, a singleton. So forth, $lfp_i(\Delta)=\{(null, c_{i-1}, c_i)\}$ for all $2\leq i\leq\pi$. Similarly, $lfp_i(\Delta)=\{(w, c_{i-1}, c_i) / w\in Q'\times Y\}$ for all $i>\pi$.
- Observe writer-reader relationships when $i>\pi$. If $\exists_{w_{w(i, n)}\neq c_{w(i, n)}}(w_{w(i, n)}, c_{i-1}, c_i)\in lfp_i(\Delta)$, then $\lambda_\Delta^0(w_{w(i, n)})=\phi$ and $c_i\notin\lambda_\Delta^{i-w(i,n)}(c_{w(i, n)})=\phi$; $(w_{w(i, n)}, c_{i-1}, c_i)$ must have been deleted from $lfp_i(\Delta)$ by Case III of no executed writer $w_{w(i, n)}$. If $\forall_{(w_{w(i, n)}, c_{i-1}, c_i)\in lfp_i(\Delta)} w_{w(i, n)}\neq c_{w(i, n)}$, then $\lambda_\Delta^{i-w(i,n)}(c_{w(i, n)})=\{c_i\}$ and $\neg(\exists c\in\lambda_\Delta^{i-w(i,n)}(c_{w(i, n)}) \wedge (c_{w(i, n)}, c_{i-1}, c)\in lfp_i(\Delta))$; $(w, p, c_{w(i, n)})$ must have been deleted from $lfp_{w(i, n)}(\Delta)$ by Case IV of no executed reader. Both lead to the contradiction that $lfp(\Delta)$ is not a least fixed point. Therefore, $w_{w(i, n)}=c_{w(i, n)}$ and $lfp_i(\Delta)=\{(c_{w(i, n)}, c_{i-1}, c_i)\}$, a singleton.

It is obtained that $\forall_{i\in[0,\,t]}\,lfp_i(\Delta)=\{(c_{w(i,\,n)},\,c_{i-1},\,c_i)\}$ and $c_0,\,c_1,\,\ldots,\,c_t$ is a $t$-computation composed from $\Delta$. Therefore, every trichoice in $lfp(\Delta)$ is executed in at least one $t$-computation composed from $\Delta$.

*Induction step* ($\|\Delta\|>0$): If $\|\Delta\|>0$, $\exists_{T\in[0,\,t]}\,|\{c_T\,/\,(w,\,p,\,c_T)\in\Delta_T\}|\geq 2$. Let **T** be the minimal time such that $\{c_T\,/\,(w,\,p,\,c_T)\in\Delta_T\} = \{\mathbf{c_{T1}},\,\mathbf{c_{T2}},\,\ldots,\,\mathbf{c_{Tm}}\}$ with $\mathbf{m}\geq 2$ executed choices. For each executed choice $c_{T_i}$, by predecessor-successor relationships, define a time series $\mathbf{\Omega(c_{T_i})} := \Omega_0(c_{T_i}),\,\Omega_1(c_{T_i}),\,\ldots,\,\Omega_t(c_{T_i})$ of transitive closures (i.e., subrelations of $\Delta_0,\,\Delta_1,\,\ldots,\,\Delta_t$ correspondingly in time), where

$\mathbf{\Omega_T(c_{T_i})} := \{(w,\,p,\,c_{T_i})\,/\,(w,\,p,\,c_{T_i})\in\Delta_T\}$ with only one executed choice $c_{T_i}$ at time T,

$\mathbf{\Omega_j(c_{T_i})} := \{(w,\,p,\,c)\,/\,(w,\,p,\,c)\in\Delta_j \wedge (w',\,c,\,c')\in\Omega_{j+1}(c_{T_i})\,\}$ in the order from time $j:=T-1$ down to 0,

$\mathbf{\Omega_j(c_{T_i})} := \{(w,\,p,\,c)\,/\,(w',\,p',\,p)\in\Omega_{j-1}(c_{T_i}) \wedge (w,\,p,\,c)\in\Delta_j\}$ in the order from time $j:=T+1$ up to $t$.

Because transitive closures $\Omega(c_{T_i})$ are obtained by predecessor-successor relationships, transitive closures $\Omega(c_{T_i})$ have their own properties (A) and (B). Because $\forall_{j\in[0,\,t]}\,\phi\neq\Omega_j(c_{T_i})\subseteq\Delta_j$ and $0=\|\Omega_T(c_{T_i})\| < \|\Delta_T\|\neq 0$, it is obtained that $\forall_{i\in[1,\,m]}\,\|\Omega(c_{T_i})\|<\|\Delta\|$. By inductive hypothesis, for all $i\in[1,\,m]$, every trichoice in $lfp(\Omega(c_{T_i}))$ is executed in at least one $t$-computation composed from subrelations $\Omega(c_{T_i})$, composed from $\Delta$, too. (1)

By properties (A) and (B), it is obtained that

$$\Delta_j = \cup_{i\in[1,\,m]}\Omega_j(c_{T_i}) \text{ and} \tag{2}$$

$$\lambda_\Delta^m(c_j) = \cup_{i\in[1,\,m]} \lambda_{\Omega(c_{T_i})}^m(c_j) \text{ for all } m\in[0,\,t-j] \text{ and all choice } c_j \text{ at time } j\in[0,\,t]. \tag{3}$$

By observing a trichoice $(c_{w(j,\,n)},\,c_{j-1},\,c_j)\in\Delta_j$ at any time $j\in[0,\,t]$ in Fig. 3, in terms of transitive closures by successive predecessor-successor relationships, i.e., by (3), it is obtained that trichoice $(c_{w(j,\,n)},\,c_{j-1},\,c_j)$ is deleted from $\Delta_j$ by Case $\boldsymbol{i}\in\{\text{I, II, III, IV}\}$
*if and only if* (4)
$(c_{w(j,\,n)},\,c_{j-1},\,c_j)$ is deleted by Case $\boldsymbol{i}$ from <u>all</u> $\Omega_j(c_{T_1}),\,\Omega_j(c_{T_2}),\,\ldots,$ and $\Omega_j(c_{T_m})$.

From (2), by (4), it is obtained that $\boldsymbol{lfp_j(\Delta) = \cup_{i\in[1,\,m]}\,lfp_j(\Omega(c_{T_i}))}$ **for all** $\boldsymbol{j\in[0,\,t]}$. (5)

By (1) and (5), every trichoice in $lfp(\Delta)$ is executed in at least one $t$-computation composed from $\Delta$, by observing the right-hand side of (5). □

**Theorem 12** (Polynomial Upper Bound Equivalence). There exist constants $k$ and $c$ such that $f(n) \leq cn^k$ for all but a finite (possibly empty) set of values of $n$, i.e., $f(n)$ is $O(n^k)$, ***if and only if*** there exists constant $k'$ such that $f(n) \leq n^{k'}+k'$ for all $n$.

*Proof.* $\Rightarrow$ Assume that there exist constants $k$ and $c$ such that $f(n) \leq cn^k$ for all but a finite (possibly empty) set of values of $n$.

Let $\mathbf{S} := \{n_1,\,n_2,\,\ldots,\,n_m\}$, the finite set of values of $n$ such that $f(n_i) > cn_i^k$ for all $i\in[1,\,m]$. Let $\boldsymbol{d}$ be the maximum value of $f(n_1),\,f(n_2),\,\ldots,\,f(n_m)$. If $S=\phi$, let $\boldsymbol{d}$ be 0. Then $f(n) \leq d+cn^k$ for all $n$.

Let $\boldsymbol{b}$ be the minimum value such that $c\leq 2^b$, then $f(n) \leq d+2^b n^k$ for all $n$.

When $n\geq 2$, $f(n) \leq d+ \boldsymbol{n^b}n^k = d+n^{k+b} \leq \underline{\boldsymbol{k+b+c+d}+n^{k+b+c+d}}$.

Observe $\underline{k+b+c+d+n^{k+b+c+d}}$ with $0^0=1$ when working with polynomial $cn^k$:
- if $n=0$, $\underline{k+b+c+d+n^{k+b+c+d}} = k+b+c+d+0 \geq f(0)$; because $d\geq f(0)$ if $0\in S$, otherwise $f(0)\leq c$.
- if $n=1$, $\underline{k+b+c+d+n^{k+b+c+d}} = k+b+c+d+1 \geq f(1)$; because $d\geq f(1)$ if $1\in S$, otherwise $f(1)\leq c$.

Therefore, $f(n) \leq \underline{k+b+c+d+n^{k+b+c+d}}$ for all $n$. Let $\boldsymbol{k'}= k+b+c+d$, then $\boldsymbol{f(n) \leq n^{k'}+k'}$ for all $n$.

⇐ Assume that there exists constant $k'$ such that $f(n) \leq n^{k'}+k'$ for all $n$.

Let **b** be the minimum value such that $k' \leq 2^b$, then $f(n) \leq n^{k'}+2^b$ for all $n$
$$\leq n^{k'}+n^b \text{ for all } n \text{ but } 0, 1 \text{ at worst}$$
$$\leq n^{k'+b}+n^{k'+b} = 2n^{k'+b}.$$

Let $c=2$ and $k = k'+b$, then $f(n) \leq cn^k$ for all $n$ but 0, 1 at worst, a finite set of values of $n$. □

Therefore, these two kinds of polynomial upper bound notations can be used interchangeably.

## Acknowledgments

Thanks to my wife for support. Thanks to all the anonymous reviewers on my early drafts, especially to the one who suggested oblivious machines. Thanks to Pieter Hofstra, Houkuan Huang, Gary Jin, Chang Su, Tong Sun, Xiaoming Sun and Xiwen Sun for their comments.

## References

[1] Aho, A. V., J. E. Hopcroft, and J. D. Ullman. *The Design and Analysis of Computer Algorithms*. Addison-Wesley, 1974.
[2] Aho, A. V. and J. D. Ullman. "Universality of data retrieval languages". In *POPL*, pages 110–119, 1979.
[3] Aaronson S. "P=?NP". In *Open Problems in Mathematics*, Springer, 2016.
[4] Codd E. F. "A relational model of data for large shared data banks". *CACM*, 13(6):377–387, 1970.
[5] Cook S. A. "The complexity of theorem proving procedures". In *STOC*, pages 151–158, 1971.
[6] Cook S. A. "The P versus NP Problem". http://www.claymath.org, 2000.
[7] Fagin R. "Generalized first-order spectra and polynomial-time recognizable sets". In *Complexity of Computation*, R. Karp (ed.), *SIAM-AMS Proceedings*, 7:43–73, 1974.
[8] Gödel K. "On formally undecidable propositions of Principia Mathematica and related systems I". 1931. In *From Frege to Gödel*, J. van Heijenoort (ed.), Harvard University Press, pages 596–617, 1967.
[9] Fortnow L. "The status of the P versus NP problem". *CACM*, 52(9):78–86, 2009.
[10] Hartmanis J. and R. E. Stearns. "On the Computational Complexity of Algorithms". *Trans. Amer. Math. Soc*. **117**:285-306, 1963.
[11] Hopcroft J. E. and J. D. Ullman. *Introduction to Automata theory, Languages, and Computation*. Addison-Wesley, 1979.
[12] Immerman N. "Relational queries computable in polynomial time". In *STOC*, pages 147–152, 1982.
[13] Karp R. M. "Reducibility among combinatorial problems". In *Complexity of Computer Computation*, R. E. Miller and J. W. Thatcher (eds.), pages 85–104, 1972.
[14] Myhill J. "Linear bounded automata". *WADD TR-60-165*, Wright Patterson AFB, Ohio, 1960.
[15] Pippenger N. and M. J. Fischer. "Relations among complexity measures". *JACM*, 26(2):361–381, 1979.
[16] R. Péter. *Recursive Functions*. Academic Press, 1956.
[17] Sipser M. "Lower bounds on the size of sweeping automata". *Journal of Computer and System Sciences*, 21:195–202, 1980.
[18] Turing A. M. "On computable numbers, with an application to the entscheidungsproblem". In *Proceedings of the London Mathematical Society*, 2(42):230–265, 1936.
[19] Vardi M. Y. "The complexity of relational query languages". In *STOC*, pages 137–146, 1982.
[20] Wigderson A. "P, NP and Mathematics - a computational complexity perspective". In *Proceedings of the ICM 06* (Madrid), vol. 1, EMS Publishing House, Zurich, pages 665–712, 2007.